\documentclass[a4paper,twocolumn,english,aps,prl,reprint,superscriptaddress,showpacs,showkeys,longbibliography, nofootinbib]{revtex4-1}
\usepackage[utf8]{inputenc}
\usepackage{amsmath,amssymb,bbm,mathrsfs,bm,braket,mathtools,color,graphicx,comment}
\usepackage{xcolor}
\usepackage{appendix}
\usepackage[colorlinks,citecolor=blue,urlcolor=blue,linkcolor = blue]{hyperref}
\usepackage[mathscr]{euscript}

\usepackage{tikz}
\usetikzlibrary{external}
\usetikzlibrary{arrows.meta}

\begin{document}

\title{Flexible constraint compilation in the parity architecture}
\author{Roeland ter Hoeven}
\affiliation{Parity Quantum Computing GmbH, A-6020 Innsbruck, Austria}
\affiliation{Institute for Theoretical Physics, University of Innsbruck, A-6020 Innsbruck, Austria}

\author{Anette Messinger}
\affiliation{Parity Quantum Computing GmbH, A-6020 Innsbruck, Austria}

\author{Wolfgang Lechner}
\email{wolfgang@parityqc.com\\wolfgang.lechner@uibk.ac.at}
\affiliation{Parity Quantum Computing GmbH, A-6020 Innsbruck, Austria}
\affiliation{Institute for Theoretical Physics, University of Innsbruck, A-6020 Innsbruck, Austria}

\newcommand{\TODO}[1]{ { \color{blue} \footnotesize (\textsf{todo}) \textsf{\textsl{#1}} } }

\date{\today}

\begin{abstract}
    We present tools and methods to generalize parity compilation to digital quantum computing devices with arbitrary connectivity graphs and construct circuit implementations for the constraint Hamiltonian of higher-order constrained binary optimization problems. In particular, we show how even non-local constraints can be efficiently implemented without expensive SWAP gates. We show how the presented tools can be used to optimize the total circuit depth and CNOT count of the quantum approximate optimization algorithm in the parity architecture and highlight the advantages of the flexible compilation using various examples. We derive the relation between the developed gate sequences and the traditional approach that uses SWAP gates. The result can be applied to improve the implementation of many other non-local operators.
    
\end{abstract}

\maketitle

\section{Introduction}
Quantum systems have improved a lot in terms of qubit numbers and coherence in the last years \cite{google_supremacy, dwave-2000, pasqal_300}, but still one of the major challenges of building scalable quantum computers is qubit connectivity. Quantum noise \cite{crosstalk, IBM_crosstalk, google_error_mitigation} such as crosstalk errors can lead to issues like frequency crowding \cite{frequency_crowding} and therefore limits the number of qubits that are connected. The parity architecture \cite{LHZ, compiler_paper, benchmark_paper, constraint_paper} offers a way to solve optimization problems, and in a recent work \cite{universal_paper} also to implement arbitrary quantum algorithms, using only local interactions.
Compiling hard optimization problems (e.g. with constraints and higher-order interactions) to the parity architecture on restricted quantum hardware is still a challenge.

In this paper, we outline a way to do parity compilation on arbitrary devices for arbitrary problems. Previous approaches required all parity constraints to be local, which made it impossible to complete the parity mapping for some sparsely connected devices (e.g. linear chains). Instead, we show how to implement non-local constraints by introducing a new concept called bridging, which allows for a more efficient implementation than using SWAP gates. The more non-local the constraints are, the more resources (e.g. CNOT gates) they require. We therefore also propose methods to minimize the non-locality of the constraints for a given optimization problem and hardware layout. 

We show that for every optimization problem mapped to the parity architecture, one can find an implementation of the constraint terms using nearest-neighbor CNOT (or other entangling) gates and local rotations. In particular, we provide tools to make an optimal choice of constraints and qubit layout to minimize the CNOT count or circuit depth.
The main steps towards this rely on the following three tasks:
\begin{itemize}
    \item[1.] Finding a valid set of constraints for a given parity mapping, where the maximum number of qubits in each constraint is restricted.
    \item[2.] Implementing (potentially non-local) constraints using local operations.
    \item[3.] Evaluating and minimizing the cost of implementing a constraint or a set of constraints based on the position of the corresponding qubits and the connectivity of the physical device.
\end{itemize}
The presented methods can be used to find a parity mapping for sparse or irregular connectivity graphs, for example on hardware platforms with missing or noisy qubits, or on platforms using non-rectangular lattice geometries \cite{heavy_hex_ibm, frequency_crowding, rigetti_octagonal}.

The next section will introduce the parity architecture and the quantum approximate optimization algorithm (QAOA) \cite{farhi2014quantum, alternating_ansatz}. After that the three main tasks defined above are addressed. Finally, we show how to solve constrained optimization problems within this framework.

 \section{The parity architecture and QAOA}
Within this work, we consider Hamiltonians defining an optimization problem of the form
\begin{align}\label{H_general}
\hat H &= \sum_{i=1}^N J_i \hat\sigma_z^{(i)} + \sum_{i=1}^N \sum_{j>i} J_{ij} \hat\sigma_z^{(i)} \hat\sigma_z^{(j)} \\ \nonumber 
&+ \sum_{i=1}^N \sum_{j>i} \sum_{k>j} J_{ijk} \hat\sigma_z^{(i)} \hat\sigma_z^{(j)} \hat\sigma_z^{(k)} + \ldots,
\end{align}
where $\hat\sigma_z^{(i)}$ are Pauli operators representing the spin variables $s_i\in\{+1, -1\}$ of the optimization problem and the $J$ coefficients define the strength of the interactions. For constrained optimization problems, there are additional side conditions
\begin{align}\label{side_conditions}
& \Big( \sum_{i=1}^N c_i \hat\sigma_z^{(i)} + \sum_{i=1}^N \sum_{j>i} c_{ij} \hat\sigma_z^{(i)} \hat\sigma_z^{(j)} \\ \nonumber 
&+ \sum_{i=1}^N \sum_{j>i} \sum_{k>j} c_{ijk} \hat\sigma_z^{(i)} \hat\sigma_z^{(j)} \hat\sigma_z^{(k)} + \ldots \Big) \ket{\psi} = C_V\ket{\psi}
\end{align}
for a state $\ket\psi$ to be a valid solution. Note that the coefficients $c$ must be consistent with the total value $C_V$ to allow for solutions. These conditions can either be added to the problem Hamiltonian by penalizing non-fulfilling configurations, or they can be fulfilled implicitly by modifying the dynamics of the optimization process such that they are never violated. We focus on the second option in this work.

Here we build on the results from Ref.~\cite{compiler_paper}, where the parity mapping was introduced for higher-order constrained binary optimization (HCBO) problems. The parity mapping  transforms the Hamiltonian by introducing a parity variable for each term (e.g. ${J_{ij} \hat\sigma_z^{(i)} \hat\sigma_z^{(j)} \xrightarrow{} J_m \hat\sigma_z^{(m)}}$) resulting in the final problem Hamiltonian:
\begin{align}\label{H_parity}
\hat H_{\rm P} &= \sum_{k=1}^K J_k \hat\sigma_z^{(k)} - \sum_{l=1}^LC_l\prod_{m_l}\hat\sigma_z^{(m_l)}.
\end{align}
Here the first term encodes the problem, originally containing $K$ interactions, as $K$ single-body terms on an increased number of spin variables. The second term compensates for the introduced redundancy by adding $L$ constraints on valid physical states $\ket\psi$
\begin{equation}\label{constraint}
    \prod_{m_l}\hat\sigma_z^{(m_l)}\ket\psi = \ket\psi,
\end{equation}
where $C_l$ has to be large enough to energetically separate the invalid states. In the rest of this paper, we will often label the physical parity qubits with the corresponding original spin variables (e.g. $J_{ij} \hat\sigma_z^{(i)} \hat\sigma_z^{(j)} \xrightarrow{} J_{ij} \hat\sigma_z^{(i, j)}$) for clarity. In the next section we will go into more detail about finding an appropriate set of constraints. 

The QAOA can then be run on the parity architecture by alternating between problem unitaries of the form 
\begin{equation}
\hat U_{\rm P} = \exp({i\gamma \hat H_{\rm P}})
\end{equation}
and driver unitaries
\begin{equation}
\hat U_{\rm X} = \exp \left({i\beta\sum_{k=1}^K\hat\sigma_x^{(k)}}\right)
\end{equation}
\cite{Lechner2018}. 
Apart from the constraints introduced by the parity mapping, all terms in $\hat U_{\rm P}$  can be implemented through single-qubit operations. In this work, we therefore focus on the efficient implementation of the multi-qubit terms appearing as constraints. We furthermore show how parity compilation can allow easy implementations of adjusted driver unitaries to ensure the satisfaction of additional constraints of the original problem (without the need to penalize them in the problem Hamiltonian \cite{alternating_ansatz}).

\section{Finding a basis of short constraints}
We define the length of a constraint as the number of qubits involved in the constraint. The longer a constraint, the more resources it will require in the implementation of the QAOA (we will look at concrete numbers in the next section). The goal is to find a basis of short constraints that allows for a valid parity mapping to the physical problem. We use the terminology of a basis of constraints equivalent to the set of constraints which generate the stabilizer of the code space (the subspace of the physical Hilbert space which corresponds to valid logical states, i.e., has no inconsistencies when mapping parity qubits to original spin variables).

Consider as an example the  problem Hamiltonian
\begin{equation} \label{example hamiltonian}
	\begin{split}
		\hat H &= J_{12} \, \hat\sigma_z^{(1)} \hat\sigma_z^{(2)} + J_{15} \, \hat\sigma_z^{(1)} \hat\sigma_z^{(5)} \\
		&+ J_{24} \, \hat\sigma_z^{(2)} \hat\sigma_z^{(4)} + J_{45} \, \hat\sigma_z^{(4)} \hat\sigma_z^{(5)} \\
		&+ J_{123} \, \hat\sigma_z^{(1)} \hat\sigma_z^{(2)} \hat\sigma_z^{(3)} + J_{345} \, \hat\sigma_z^{(3)} \hat\sigma_z^{(4)} \hat\sigma_z^{(5)}.
	\end{split}
\end{equation}
The generator matrix, which maps the original problem variables into the code subspace \cite{compiler_paper}, can be constructed by writing the terms as columns in a matrix. Each column will have a $1$ in row $i$ if $\hat\sigma_z^{(i)}$ is in the corresponding term and $0$ otherwise. In the example, $\hat\sigma_z^{(1)} \hat\sigma_z^{(2)}$ becomes the first column, with non-zero entries in the first and second row). We thus arrive at the generator matrix

\begin{equation} \label{example generator matrix}
	\mathbf{G} = \begin{pmatrix}
		1 & 1 & 0 & 0 & 1 & 0 \\
		1 & 0 & 1 & 0 & 1 & 0 \\
		0 & 0 & 0 & 0 & 1 & 1 \\
		0 & 0 & 1 & 1 & 0 & 1 \\
		0 & 1 & 0 & 1 & 0 & 1
	\end{pmatrix}.
\end{equation}

The parity check matrix $\mathbf{P}$ can then be found by solving the equation 
\begin{equation}
\mathbf{G} \mathbf{P}^\top \equiv 0\mod 2
\end{equation}
for a matrix of maximal rank. For the generator matrix in Eq.~\eqref{example generator matrix}, a possible solution for the parity check matrix is

\begin{equation} \label{example parity matrix}
	\mathbf{P} = \begin{pmatrix}
		1 & 1 & 1 & 1 & 0 & 0 \\
		1 & 0 & 0 & 1 & 1 & 1
	\end{pmatrix}.
\end{equation}

The rows in this matrix represent a basis of the constraint space to be implemented; we call this the target constraint space. Similar to the generator matrix, a $1$ in column $i$ of the parity check matrix indicates that there is a term $\hat\sigma_z^{(i)}$ in the corresponding constraint. Any basis of the constraint space describes a valid choice for a set of constraints to be implemented in the mapped problem Hamiltonian.
However, we can find many different choices for a basis by performing elementary row operations. Note that, as the entries of $\mathbf{G}$ and $\mathbf{P}$ are in $\mathbb{Z}_2$, all operations are performed modulo 2. The length of a constraint is its number of non-zero entries. In general, a constraint can be as long as the number of terms in the original Hamiltonian and it is a hard problem to find a basis that only consists of constraints up to a certain length. However, if we are only interested in finding a basis of short constraints, for example of length $3$ and $4$, this can be done efficiently. For a fixed number $L$, all constraints of length $l\leq L$ in a system with $K$ Hamiltonian terms can be found in complexity $\mathcal{O}(K^L)$.

The simplest way to achieve this is by taking all combinations of $L$ physical spin variables and checking if their product forms a valid constraint. We can confirm that a constraint is valid if the corresponding row vector is in the target constraint space. Alternatively, we can check that each logical index occurs an even amount of times in the product. If that is the case, the product of physical spin variables in the constraint is $1$ by definition. For example, the qubit product 
\begin{equation}
    \hat\sigma_z^{(1, 5)} \hat\sigma_z^{(2, 4)}\hat\sigma_z^{(1, 2, 3)}\hat\sigma_z^{(3, 4, 5)} 
\end{equation}
forms a valid constraint as
\begin{equation}
    \hat\sigma_z^{(1)} \hat\sigma_z^{(5)} \hat\sigma_z^{(2)} \hat\sigma_z^{(4)}
\hat\sigma_z^{(1)} \hat\sigma_z^{(2)} \hat\sigma_z^{(3)} \hat\sigma_z^{(3)} \hat\sigma_z^{(4)} \hat\sigma_z^{(5)} =\hat I.
\end{equation}
This can equivalently be verified by checking that it is in the target constraint space directly, using the found basis and Gaussian elimination.

We can then grow a basis of short constraints, by adding to it each time we find a new and linearly independent combination of qubits that form a valid constraint. The process is finished when the constraints in the basis span the target constraint space. However, it is possible that the short constraints are not enough to span the target constraint space, in which case we can use ancilla qubits to break down long constraints into short ones. Consider the following parity check matrix, which emerges in the parity mapping of a logical problem graph that is a single cycle of length $5$:

\begin{equation} \label{long parity matrix}
	\mathbf{P} = \begin{pmatrix}
		1 & 1 & 1 & 1 & 1 & \\
	\end{pmatrix}.
\end{equation}
It consists of a single constraint of length $5$. To obtain a basis of constraints of length $3$ and $4$, we need to add an ancilla and a row to this parity check matrix, e.g.,
\begin{equation} \label{broken up parity matrix}
	\mathbf{P} = \begin{pmatrix}
		1 & 1 & 1 & 0 & 0 & 1\\
  		0 & 0 & 0 & 1 & 1 & 1\\
	\end{pmatrix}.
\end{equation}
Here, the last column corresponds to the added ancilla. The original constraint can be found by adding the first and second row and eliminating the ancilla. With this process, any long constraint can be reduced to a number of small constraints using some ancilla qubits.

In summary, the process for a general problem will have the following steps.
First, enumerate all combinations of $3$ and $4$ qubits and find all valid constraints that can be made. If these constraints span the target constraint space, pick a smallest subset of them which still spans the target constraint space. Otherwise, if there are still missing constraints, go over the original basis of (potentially long) constraints of the target constraint space and identify the constraints which are not in the initial space of short constraints. For these constraints, it is impossible to find a decomposition into short constraints, so we need ancillas to break them down. Add the resulting constraints including ancillas to the basis of the short constraints and keep iterating over the target constraint space basis until all constraints are included. 

Note that in general, it is not necessary to stick to a basis of constraints of lengths $3$ and $4$, because constraints of arbitrary length can be implemented. However, allowing longer constraints increases the search space and can also cost more CNOT gates. In some cases it is possible to avoid using ancillas by allowing longer constraints, e.g., for Eq.~\eqref{long parity matrix} no ancillas are needed when constraints of length $5$ are directly implemented. There can also be a trade-off between necessary number of ancillas and circuit depth, as long constraints take more time to execute and are harder to parallelize. 

\section{Gate sequences for constraints}
Consider a quantum device that can do single-qubit operations on all qubits and an entangling 2-qubit gate between some of the qubits such that a universal gate set over all qubits is achieved. The connectivity graph $G$ of the device, where all qubits are nodes and the qubits that can directly interact with each other have an edge, is then a connected graph. 

In the following, we show how to efficiently implement a constraint $C$ with the operator 
\begin{equation}\label{constraint_operator}
\hat{U}_C = \exp \left(i \alpha_{C} \prod_{m \in C} \hat\sigma_z^{m}\right)
\end{equation}
in the QAOA for any arrangement of the qubits in $C$ and any angle $\alpha_C$ using a decomposition into CNOT gates and local rotations. 
Any $n$-qubit rotation $\exp(i\alpha \hat\sigma_z^{\otimes n})$ can be decomposed into an $n-1$-qubit rotation $\exp(i\alpha \hat\sigma_z^{\otimes n-1})$ and two CNOT gates with control on the $n$-th qubit and target on any of the $n-1$ other qubits, applied before and after the rotation (the target must be the same qubit for both CNOT gates) \cite{plaquette_optimizer}. This can be iteratively applied to decompose any many-body rotation, such as the constraint operator in Eq.~\eqref{constraint_operator}, into a sequence of CNOT gates followed by a single-body rotation and the reverse sequence of CNOT gates. 

The effect of such CNOT sequences can be understood by looking at the effect of a single CNOT gate, collecting the parity of its control and target qubit at the target,
\begin{equation}
    \ket{a}\ket{b} \overset{{\rm CNOT}}{\longmapsto} \ket{a}\ket{a\oplus b},\,\,\,\,\,a,b\in{0,1}.
\end{equation}
Both control and target qubits can already hold the parity of other states, for example if they have been targeted by other CNOT gates before. A $Z$-rotation on a physical qubit which holds the parity of multiple (logical) qubit states then effectively performs a collective rotation on all these logical qubits by adding a phase only to those states whose corresponding multi-qubit parity is odd. After having performed such a rotation, the original CNOT sequence can be applied in reverse order to return to the original representation of qubits.

For a given constraint $C$, there are many different options for such sequences, as any CNOT sequence which collects the parity of the qubits in $C$ towards a single qubit can be used. The effect on any other qubits is irrelevant as it is undone afterwards by the reversed CNOT sequence. However, as the connectivity of most quantum devices is limited, we are interested in a sequence using only local interactions. 
We will show that for any tree in the connectivity graph $G$ which spans the qubits in $C$, we can find such a sequence using CNOT gates along the edges of that tree. We call such a tree the Steiner tree of $C$ in $G$. The resulting decomposition has depth $l_{\rm max} + \mathcal{O}(1)$, where $l_{\rm max}$ is the largest distance (i.e., the maximal number of edges) between two constraint qubits in the tree.

\subsection{Local constraints}
\begin{figure}
    \centering
    \includegraphics[width=\columnwidth]{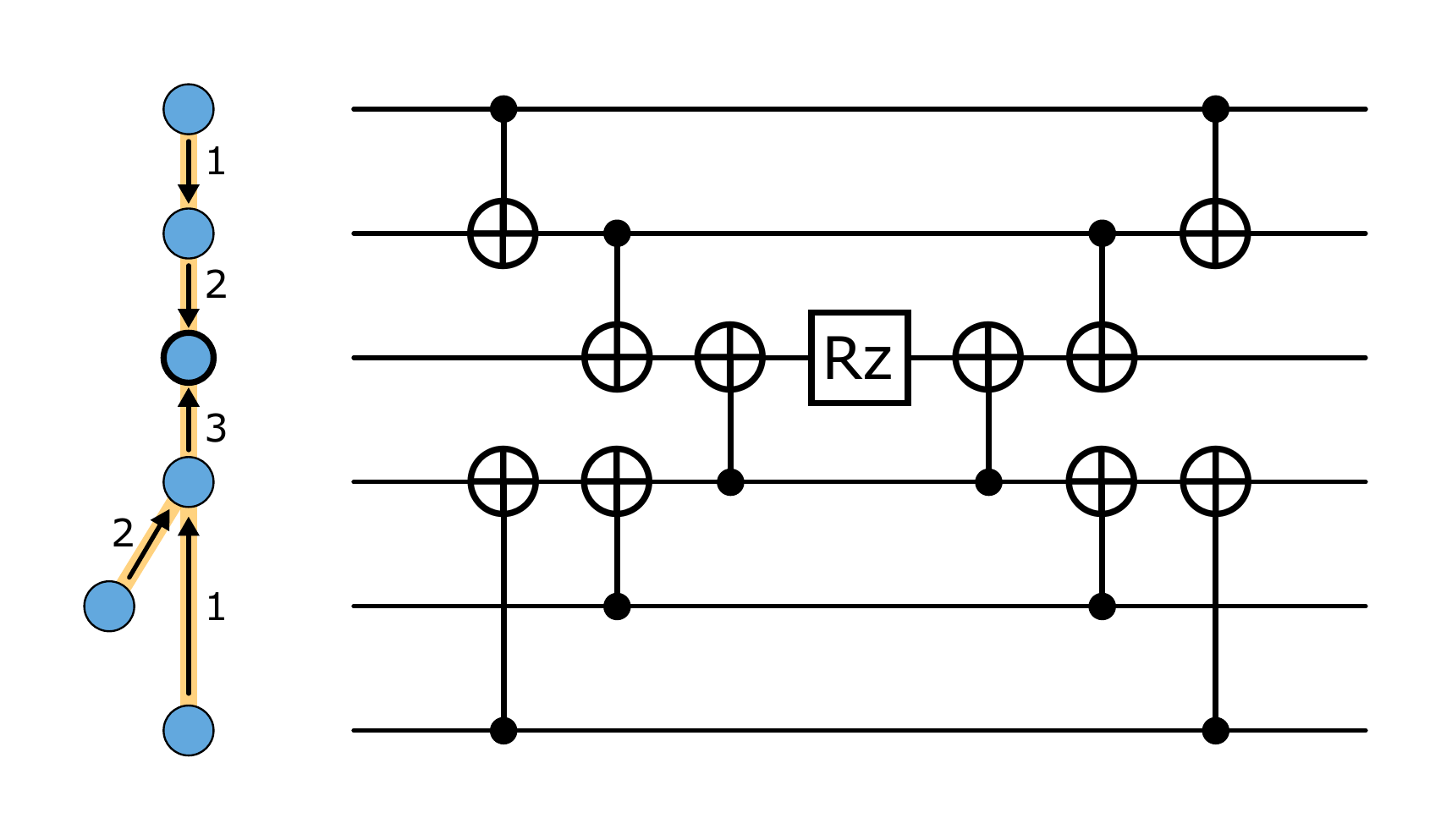}
    \caption{Implementation of a local constraint on six qubits using the Steiner tree highlighted in yellow. Numbers and arrows indicate the time step and direction (control to target) of a CNOT gate on the given edge before the rotation gate at the root. After the rotation, the CNOT gates are implemented with the same direction, but in opposite order.}
    \label{fig:local_constraint}
\end{figure}

If there is a Steiner tree of a constraint $C$ in the connectivity graph $G$ which only contains the qubits in $C$, we call this constraint local.
For a local constraint, we can find an implementation as follows.
First, we choose a root of the tree on which we want to perform the single-body rotation. Then we can obtain the CNOT sequence to be applied after the rotation by traversing the tree from the chosen root towards the leaves and adding a CNOT gate at every edge, controlling the child (outer) node and targeting the parent (inner) node. The CNOT sequence to be applied before the rotation is then the inverse of this. The resulting decomposition is shown for an example tree in Fig.~\ref{fig:local_constraint}. The root of the tree can be any qubit in $C$ but is usually chosen such that its longest distance to any leaf is minimized, which typically corresponds to the implementation with the smallest circuit depth\footnote{Note that on trees with many branches, the circuit depth does not only depend on the distance between the root and the leaves but also on the order of the nodes in the tree.}. We furthermore have the freedom to choose the order of CNOT gates which are on edges entering the same node (i.e., target the same qubit). Both the choice of the root and the order of CNOT gates with the same target can be used to further optimize the resulting circuit depth, either of the individual constraint, or of a larger circuit implementing multiple constraint operators in parallel.

\subsection{Non-local constraints}
\begin{figure}
    \centering
    \includegraphics[width=\columnwidth]{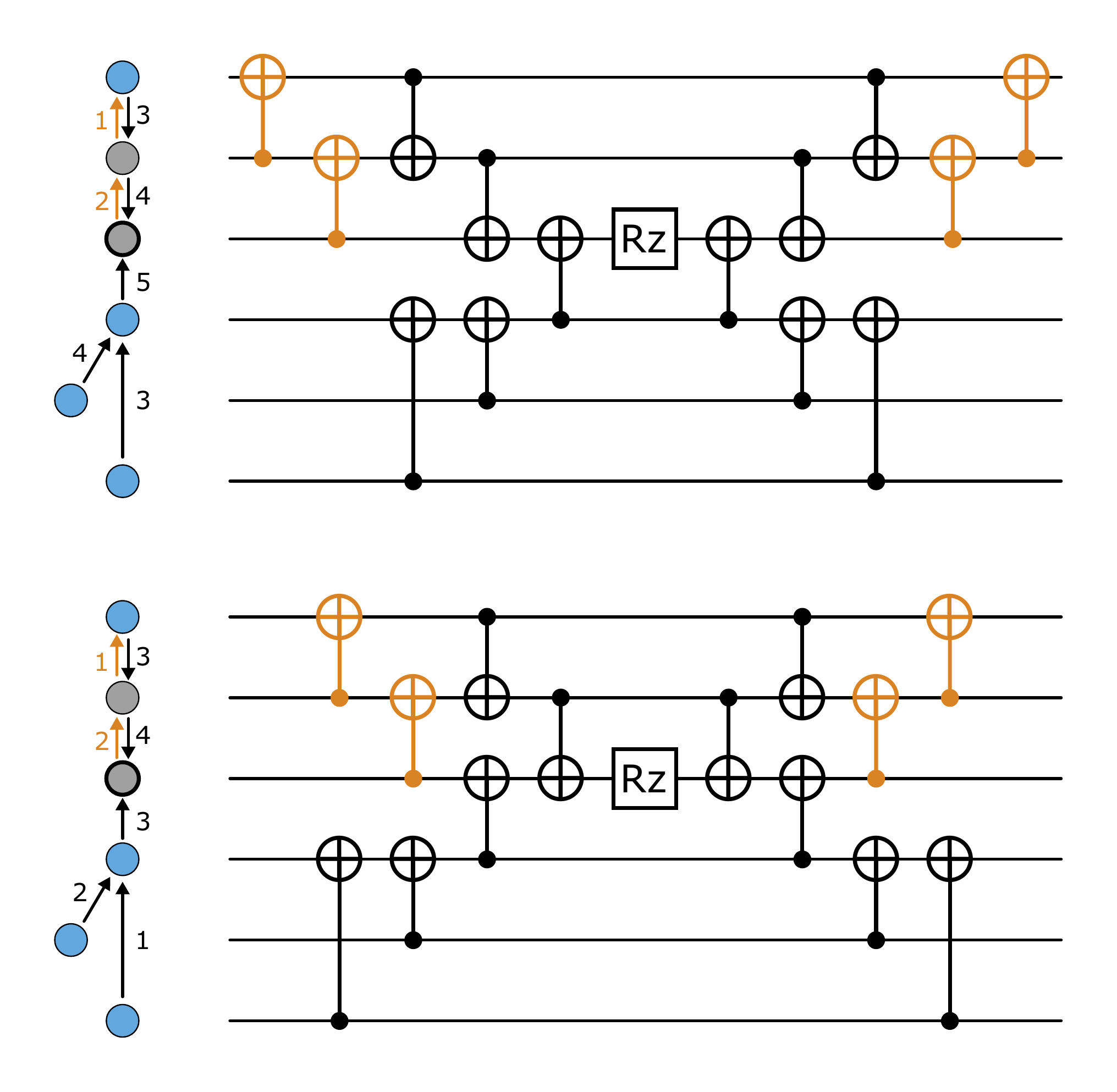}
    \caption{Implementation of a non-local constraint on four qubits (blue) with two other qubits (gray) along the Steiner tree in the same representation as Fig.~\ref{fig:local_constraint}. Orange CNOT gates indicate additional gates that are required to cancel out the effect of the unwanted qubits along the tree. The top circuit shows the result when the optimal-depth CNOT sequence for the full tree is simply extended by the additional gates. In the bottom circuit, the CNOT sequence of the full tree has been rearranged to further reduce the depth of the full circuit.}
    \label{fig:nonlocal_constraint}
\end{figure}

If the Steiner tree of a constraint $C$ contains qubits which are not in $C$, one can use a similar strategy to implement the constraint. Instead of using SWAP gates to bring the constraint qubits next to each other, we derive a sequence of CNOT gates to effectively bridge the non-constraint qubits. We engineer the CNOT sequence such that every constraint qubit appears once in the final parity of the rotation qubit, but every bridged qubit along the way appears twice and thus its effect cancels out. For this, we start with the CNOT sequence which implements a constraint on the full tree, including the non-constraint qubits along the way. We then add CNOT gates in the beginning and end of the constraint circuit, which add the information of each non-constraint qubit as a parity onto exactly one other qubit, as for example shown in Fig.~\ref{fig:nonlocal_constraint}. 

\begin{figure}
    \centering
    \includegraphics[width=.9\columnwidth]{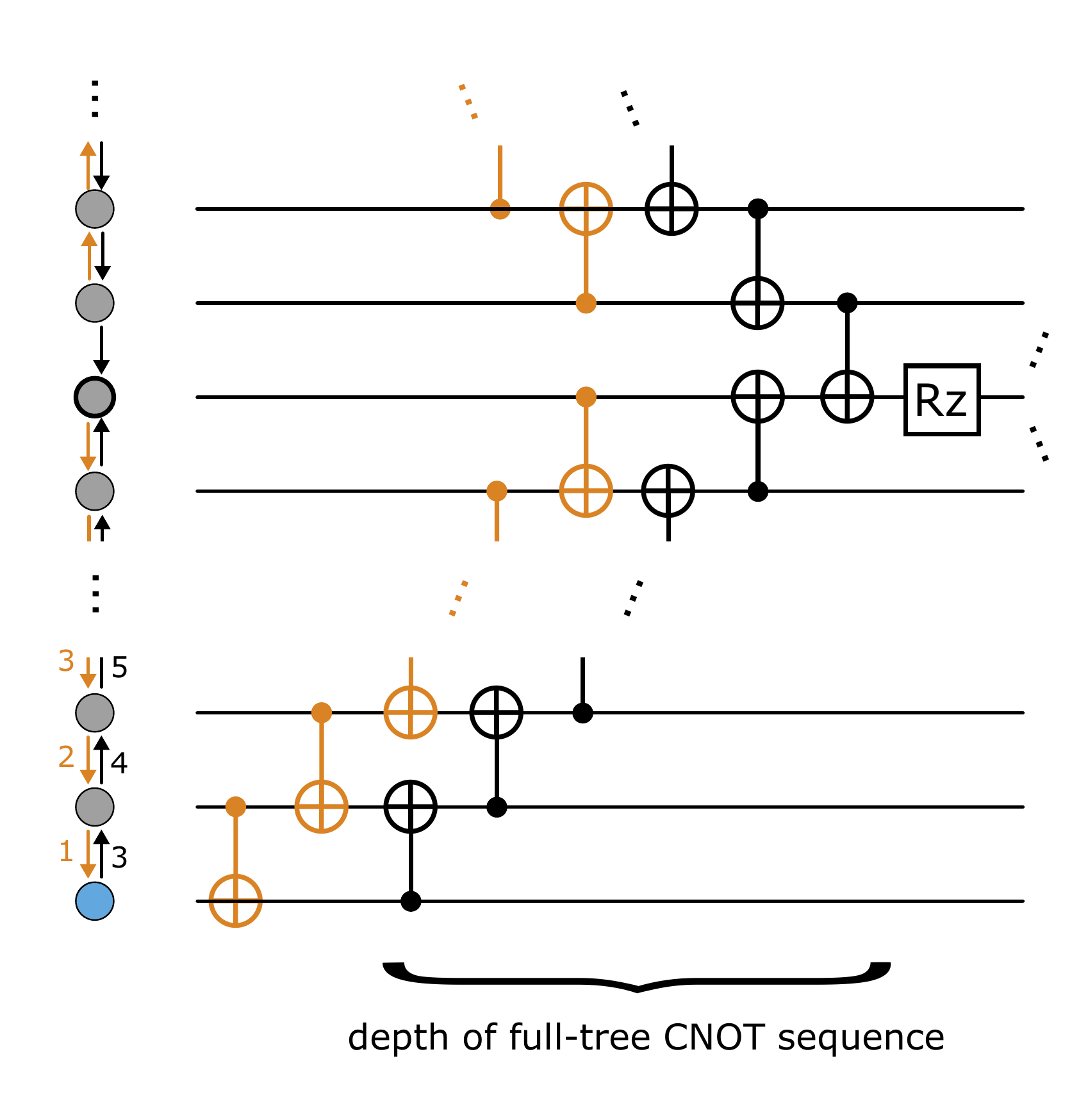}
    \caption{Illustration of the worst-case circuit depth for non-local constraints. On any branch of the Steiner tree, the CNOT gates required for bridging non-constraint qubits will protrude from the CNOT sequence implementing the full Steiner tree by at most two time steps. As the same sequence is repeated in reverse after the rotation in the center (not shown in this figure), the maximal depth increase compared to a constraint on the full Steiner tree is four.}
    \label{fig:large_constraint}
\end{figure}

Together with the CNOT sequence over the full Steiner tree $T$, the information of those qubits is collected twice, once from the original qubit and once from the qubit which was targeted with the additional CNOT gate. The final operation which is performed is then
\begin{align}
\hat{U}_C &= \exp \left(i \alpha_{C}
\prod_{m \in T} \hat\sigma_z^{m} 
\prod_{m \in T \setminus C} \hat\sigma_z^{m} \right)\\
&=\exp \left(i \alpha_{C}
\prod_{m \in C} \hat\sigma_z^{m} \right),
\end{align}
where $T \setminus C$ is the set of qubits which are in the Steiner tree but not in the constraint.
Whenever $T \setminus C$ contains more than one qubit, the order of the additional CNOT gates has to be chosen such that they don't duplicate qubit information more than once. Consider the case of bridging two neighboring qubits $i$ and $j$ along a tree. A gate $\text{CNOT}_{ij}$ followed by $\text{CNOT}_{jk}$, for example, would transform a basis state $\ket{a_i}\ket{a_j}\ket{a_k}$ to the undesired state $\ket{a_i}\ket{a_i \oplus a_j}\ket{a_i \oplus a_j \oplus a_k}$, where qubit $a_i$ now appears three times. The reverse sequence, however, results in the desired state $\ket{a_i}\ket{a_i \oplus a_j}\ket{a_j \oplus a_k}$, duplicating exactly $a_i$ and $a_j$.
Besides the order of CNOT gates, we also need to choose the target qubit, i.e., on which qubit we duplicate the information. Any qubit neighboring the unwanted qubit in the Steiner tree $T$ is a valid choice for this, no matter if it is a constraint qubit or not.
As we want to construct a circuit with minimal depth, we choose to duplicate the information of every unwanted qubit to its child node in the tree, i.e., with a CNOT in the opposite direction of the following full Steiner tree implementation. This way, the additional CNOT gates can be applied almost in parallel to the Steiner tree implementation and result in a maximal depth increase of four CNOT steps compared to the full Steiner tree implementation (see Fig.~\ref{fig:large_constraint}).

The total number of CNOT gates required for a constraint $C$ with Steiner tree $T$ can then be calculated from the gates required for the full Steiner tree implementation and the additional CNOT gates for bridging unwanted qubits as:
\begin{equation}
n_\text{CNOT} = 2(|T| - 1) + 2(|T| - |C|) = 4|T| - 2|C| - 2,
\end{equation}
where $|\cdot|$ is the number of qubits in the tree or the constraint. 

\begin{figure}
    \centering
    \includegraphics[width=\columnwidth]{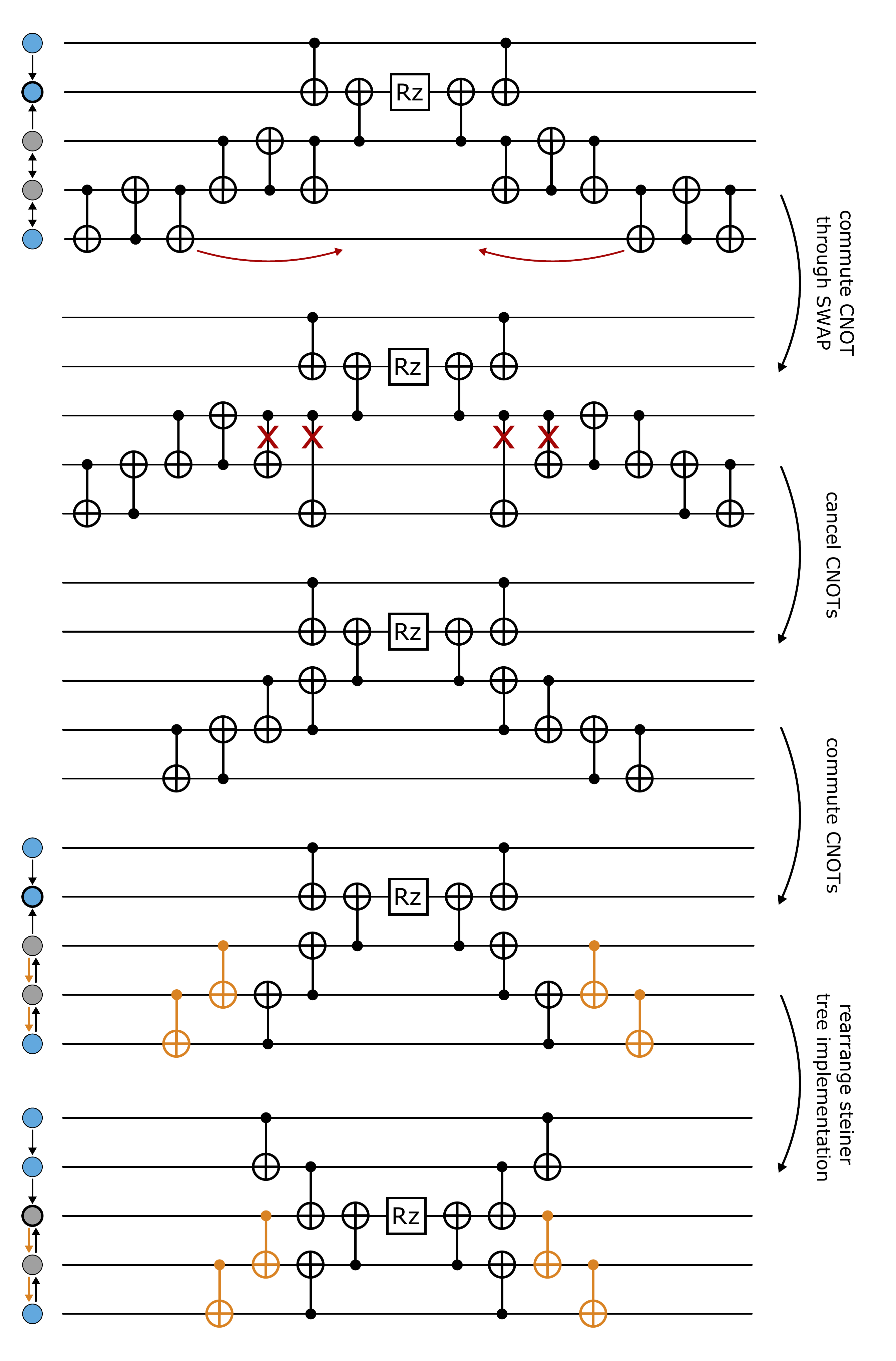}
    \caption{Derivation of a bridging sequence of a non-local constraint operator $\exp(i \alpha \hat\sigma_z\hat\sigma_z\hat\sigma_z)$ on the qubits highlighted in blue from an equivalent implementation via SWAP gates. The same derivation also works for other operators as long as there exists a SWAP decomposition whose last gate commutes with the target operator.}
    \label{fig:swap_to_bridge}
\end{figure}

\subsection{Circuit equivalence to SWAP sequence}
The presented gate sequence is equivalent in its effect to the conventional method of rearranging the constraint qubits via SWAP gates until the constraint is local. Figure~\ref{fig:swap_to_bridge} shows how an efficient bridge sequence can be derived from an approach using SWAP gates on a simple example. The derivation also works for more complicated constraint arrangements, and in general can also be applied to other non-local interactions whenever the interaction commutes with $\hat\sigma_x$ or $\hat\sigma_z$ on each of its qubits. Even if the interaction commutes with neither of them, the procedure can still work if the SWAP gate is decomposed into other gates (e.g.,~controlled-$Y$ gates if the non-local interaction commutes with $\hat\sigma_y$). The main step in the simplification of the SWAP sequence lies in cancellation of gates which appear again in the reverse SWAP sequence and commute with everything in-between. If the implemented interaction commutes with $\hat\sigma_z$ (and thus with the controls of CNOT gates), the SWAP gates must be decomposed such that they begin and end with a CNOT gate controlling the qubits on the inside (where the desired interaction is implemented). Similarly, if the interaction commutes with another operator, the SWAP gates must be decomposed accordingly. 
CNOT gates resulting from a SWAP decomposition further away can still be commuted towards the inside by taking into account the other SWAP gates on the way (see first step in Fig.~\ref{fig:swap_to_bridge}). With this, the total cost can be reduced from three to two gates per SWAP gate. Finally, rearranging the gates, taking into account the different possibilities to implement constraint operators, can further improve the circuit depth.
For longer constraints along more complex trees, the same idea still holds, although becoming less straightforward.

\subsection{Scaling comparison}
To compare our solution to an implementation using SWAP gates, we consider a simple constraint of two qubits of distance $l$ along the connectivity graph. Such a constraint requires
\begin{equation}\label{num_cnots_bridge}
n_\text{CNOT}^\text{(bridge)}=2+4(l-1)
\end{equation}
CNOT gates when using the bridging method, whereas it would take  
\begin{equation}
n_\text{CNOT}^\text{(SWAP)}=2+6(l-1)
\end{equation}
when using SWAP gates (implemented by three CNOT gates per SWAP gate). The corresponding circuit depths of this constraint implementation are
\begin{equation}
d^\text{(bridge)} \leq 2\left\lceil \frac{l+1}{2} \right\rceil + 4,
\end{equation}
where the linear term is the depth of implementing a constraint operator on the full steiner tree, and the constant is due to the additional bridging gates, and
\begin{equation}
d^\text{(SWAP)}=6\left\lceil \frac{l+1}{2} \right\rceil -4
\end{equation}
when assuming a SWAP sequence without further optimization. Note that the constant overhead of four additional steps in the bridge implementation does not apply for small $l$, where either no bridging is necessary ($l=1$), or only two additional steps are necessary ($1<l<5$).
While existing transpilers \cite{Qiskit, pytket} are likely able to cancel some of the CNOT gates of the occurring SWAP sequences, we stress that the main advantage of the construction via bridging lies in the rearrangement of the gates into a minimal-depth circuit which is not always straightforward using local circuit optimization strategies. The construction via parity, however, provides simple rules and allows flexibility for optimization.
For longer constraints, the scaling of the circuit depth mainly depends on the largest distance of two constraint qubits in the Steiner tree, while the gate count is determined by the total length of the tree.

\section{Minimal Steiner trees for non-local constraints}
Given that the cost of implementing constraint operators (both gate count and depth) scale linearly with the size of the corresponding Steiner tree, we now want to find the smallest such Steiner tree. The minimal tree that spans a subset of vertices (terminals) of the connectivity graph is well-known from graph theory and called the minimal Steiner tree. We define the size of a tree as its number of edges. 

There are many (approximate) algorithms to find minimal Steiner trees \cite{approx_steiner_tree} in arbitrary graphs and also more specific algorithms that use the geometry of the graph.

For rectangular devices with nearest-neighbor connections, we get a rectilinear Steiner tree problem, first studied by Hanan \cite{Hanan}. He introduced the Hanan grid, which is obtained by drawing horizontal and vertical lines through all terminals. There always exists an minimal Steiner tree on the Hanan grid \cite{hanan_proof}. The rectilinear Steiner tree problem is also well-studied in the presence of obstacles \cite{obstacle_rect_steiner}, which is relevant because realistic devices may have obstacles like defect qubits and missing qubit connections. 

\begin{figure}
    \includegraphics[width=\columnwidth]{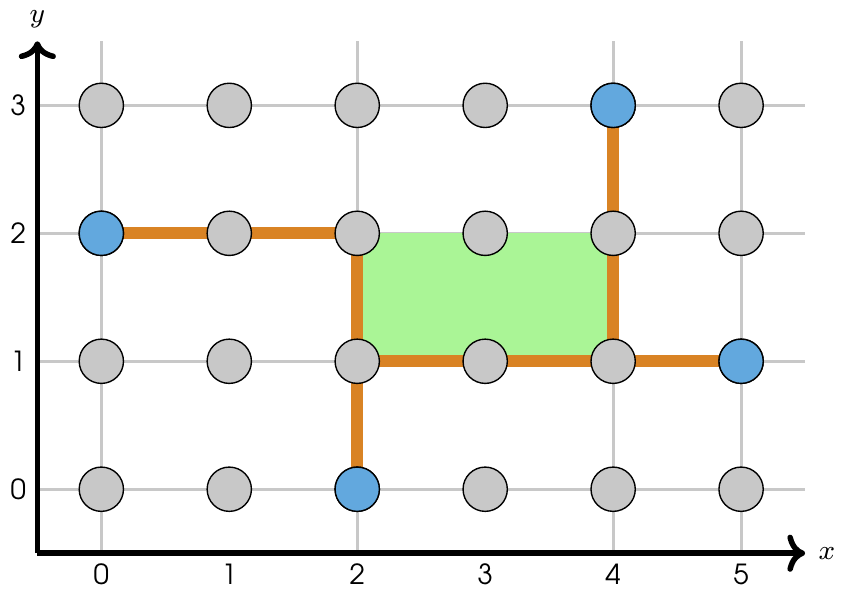}
    \centering
    \caption{Example of a minimal Steiner tree for the $4$ blue nodes (terminals) on a rectangular lattice. There are many minimum Steiner trees of size $9$, one of which is drawn in orange. The grey lines represent the Hanan grid, constructed by drawing horizontal and vertical lines through all terminals. There always exists a minimal Steiner tree on the Hanan grid for any choice of terminals. The central rectangle that can be used to find optimal Steiner trees for $4$ terminals is highlighted in green.}
    \label{fig:steiner_tree}
\end{figure}

For short constraints up to a certain size (number of constraint qubits, independent of their position or distance in the connectivity graph), there is a fast way to find rectilinear Steiner trees. Here, we will show how to efficiently find minimal Steiner trees for constraints of length $3$ and $4$. The minimal Steiner tree can be found by sorting the terminals in the $x$ direction and in the $y$ direction and only keeping the component that was sorted for, such that $x_1 \leq x_2 \leq x_3 (\leq x_4)$ and $y_1 \leq y_2 \leq y_3 (\leq y_4)$ for 3-qubit (4-qubit) constraints. For 3-qubit constraints, the smallest tree size $S_3$ is
\begin{equation}\label{steiner_tree_3}
    S_3 = (x_3 - x_1) + (y_3 - y_1),
\end{equation}
which is the largest difference between vertices in the horizontal and vertical direction. For 4-qubit constraints (see Fig.~\ref{fig:steiner_tree}), a minimal Steiner tree can be found by first connecting all terminals to the closest corner of the central rectangle defined by its bottom left corner $(x_2, y_2)$ and top right corner $(x_3, y_3)$, and then connecting the corners of the central rectangle. If the terminals connect to all four corners, this requires one connection on the longer side and two connections on the shorter side, otherwise a single connection in each direction suffices. The size of the minimal Steiner tree $S_4$ is then
\begin{equation}\label{steiner_tree_4}
    S_4 \leq (x_4 - x_1) + (y_4 - y_1) + \min(y_3 - y_2, x_3 - x_2),
\end{equation}
where the additional term is the length of the shorter side of the central rectangle, which can contribute twice to the total length.

In Fig.~\ref{fig:steiner_tree} we have $(x_1, x_2, x_3, x_4) = (0, 2, 4, 5)$ and $(y_1, y_2, y_3, y_4) = (0, 1, 2, 3)$, which results in the tree size $S_4 = 5 + 3 + \min(1, 2) = 9$. 

\section{Application: 1D connectivity}
In this section, we look at a small example and compare to a direct implementation using the standard gate model. The Hamiltonian we will look at has the form
\begin{equation}\label{eq:comparison_hamiltonian}
	\begin{split}
		\hat H &= 
        J_{12} \, \hat\sigma_z^{(1)} \hat\sigma_z^{(2)} + 
        J_{13} \, \hat\sigma_z^{(1)} \hat\sigma_z^{(3)} + 
        J_{14} \, \hat\sigma_z^{(1)} \hat\sigma_z^{(4)} \\ &+ 
        J_{23} \, \hat\sigma_z^{(2)} \hat\sigma_z^{(3)} + 
        J_{123} \, \hat\sigma_z^{(1)} \hat\sigma_z^{(2)} \hat\sigma_z^{(3)} + 
        J_{124} \, \hat\sigma_z^{(1)} \hat\sigma_z^{(2)} \hat\sigma_z^{(4)} \\ &+ 
        J_{134} \, \hat\sigma_z^{(1)} \hat\sigma_z^{(3)} \hat\sigma_z^{(4)} + 
        J_{234} \, \hat\sigma_z^{(2)} \hat\sigma_z^{(3)} \hat\sigma_z^{(4)}
	\end{split}
\end{equation}
and the goal is to compile this problem on a 1-dimensional device with nearest-neighbor interactions only. Under these restrictions, it is impossible to compile with only local constraints, so we need bridging to successfully compile it. We will then compare the gate count and circuit depth to a standard implementation of the gate model. 

\begin{figure}
    \includegraphics[width=\columnwidth]{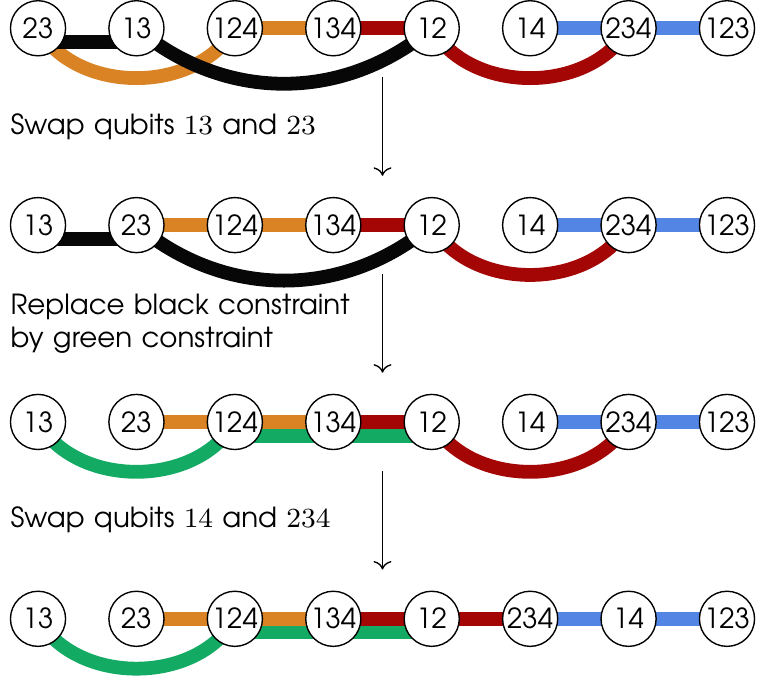}
    \centering
    \caption{Process of parity compilation with bridging, optimizing gate count and depth. Qubits are marked by their logical indices (e.g., $23$ means that a $\hat\sigma_z$ on this qubit corresponds to the logical operation $\hat\sigma_z^{(2)} \hat\sigma_z^{(3)}$) and each colored line connects the qubits of a parity constraint. The two possibilities for optimization are highlighted: a qubit swap and a change of the constraint basis. In each step of the process, the constraints become more local and thus cheaper to implement. In the first step the orange constraint is made local by swapping qubits $13$ and $23$. Then the black constraint is replaced by the product of itself with the orange constraint, resulting in the constraint shown in green. The green constraint is slightly cheaper because it only has to bridge a single qubit. Finally the red constraint is made local by swapping qubits $14$ and $234$.}
    \label{fig:1d_example}
\end{figure}

In general, any choice of the constraint basis and layout of qubits can be implemented using bridging. However, it is possible to optimize the choice of constraint basis and qubit layout to obtain constraints that are as local as possible. Both the constraint basis and qubit layout can be incrementally changed to improve the quality, as for example shown in Fig.~\ref{fig:1d_example}. Optimization can not only be done for individual constraints, but also to parallelize multiple constraints, which requires a minimization of the constraint overlap during optimization. Often, these two considerations agree, which means that making individual constraints as local as possible will also result in constraints that have small overlap and are therefore easier to parallelize.

The required amount of gates to implement the three local 3-qubit constraints and the non-local 4-qubit constraint in the last line of Fig.~\ref{fig:1d_example} can now be calculated. A local 3-qubit constraint requires $4$ CNOT gates, so the three constraints require $3 \cdot 4 = 12$ CNOT gates. The 4-qubit constraint needs to bridge one qubit and requires $10$ CNOT gates. This means that in total we need $22$ CNOT gates to implement the entire circuit. 

For a standard implementation for the gate model on a 1-dimensional device, a possible good layout is:
\begin{equation}\label{eq:gm_layout}
    4 -- 1 -- 2 -- 3.
\end{equation}
No matter how the qubits $1$, $2$, $3$ and $4$ are arranged, there will always be two local 3-qubit terms and two non-qubit 3-body terms. The local 3-qubit terms require $4$ CNOT gates and the non-local terms $8$ CNOT gates. So in total the 3-qubit terms require $2 \cdot 4 + 2\cdot 8 = 24$ CNOT gates.
For the 2-qubit interactions, only $3$ of them can be local and the last one must be non local. In Eq.~\eqref{eq:gm_layout} we have $3$ local 2-qubit terms and $1$ term which has to bridge a qubit. For the local terms we require $2$ CNOT gates and for the last term $6$ CNOT gates, which means that in total we require $3 \cdot 2 + 6 = 12$ CNOT gates for the 2-qubit terms\footnote{Note that this already assumes further optimization of the circuit; a naive implementation with swap gates would take $14$ CNOT gates.}. Overall, the standard gate model implementation requires $24 + 12 = 36$ CNOT gates. 

In summary, this example shows that bridging allows us to use parity compilation even on sparsely connected devices and can give an advantage over the standard gate model implementation. Especially problems with higher-order interactions are natural to solve with the parity approach, as recent benchmarks indicate \cite{benchmark_paper}. 

\section{Application: Constrained optimization problems}
The parity architecture allows one to solve different types of constrained optimization problems without overhead. 
In Ref.~\cite{compiler_paper} it was shown that problem constraints of product form (e.g. $\hat\sigma_z^{(3)} \hat\sigma_z^{(6)}=\hat I$) can directly be included when calculating the parity check matrix. Since each product constraint represents a single parity qubit that is fixed to a certain value, this parity qubit can be left out and implicitly implemented using the parity constraints. These product constraints can therefore lead to a reduction in qubit and gate count, whereas in other implementations they can lead to further overhead.
In Ref.~\cite{constraint_paper}, more general (polynomial) problem constraints are investigated. The main idea there is to use the Hamiltonian dynamics of the driver term to satisfy the problem constraints. In the initial state the spins are put in a state that satisfies the constraint and the driver only allows a specific kind of exchange between spins, which makes sure that all constraints stay satisfied during the algorithm. For all parity qubits which are part of such a polynomial constraint, the corresponding single-body $\hat\sigma_x$ term is removed from the driver Hamiltonian, and instead an exchange Hamiltonian of the form
\begin{equation}
    \label{eq:Hexchange}
	H_{\textrm{exch.}} = \sum_{\langle i,j \rangle} \tilde{\sigma}^{(i)}_+ \tilde{\sigma}^{(j)}_-  +  h.c.
\end{equation}
is added, where $\langle i, j \rangle$ are pairs of neighboring parity qubits that span the polynomial constraint. Such an exchange interaction is available on different hardware platforms like for example neutral atoms \cite{exchange_rydberg, exchange_rydbergZoller} or superconducting qubits \cite{exchange_transmon, exchange_coupled_transmon, exchange_supercond_martines}.  However, since it is typically only available locally, the qubits forming a polynomial constraint must be placed directly next to each other on the physical hardware. This sets a limit to the flexibility of qubit placement. Apart from the placement restrictions, the compilation process is not affected by polynomial constraints. The parity check matrix is independent of these constraints, as they are enforced exclusively with the driver Hamiltonian. 

\begin{figure}
    \includegraphics[width=0.8\columnwidth]{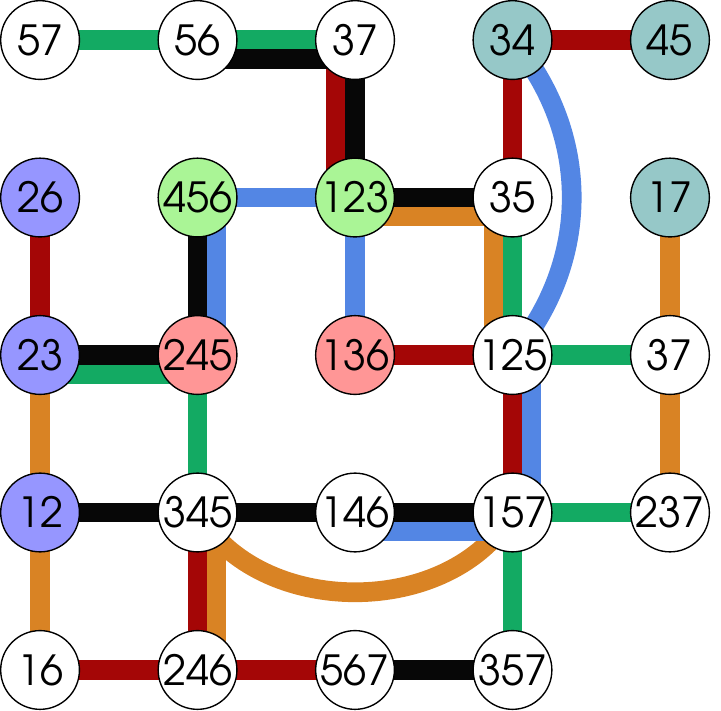}
    \centering
    \caption{Compilation of a constrained optimization problem with $2$ product constraints and $4$ general polynomial constraints, requiring connectivity only on a subset of the edges of a square lattice. The colored qubits are in polynomial constraints, where each of the four colors represents a different constraint, e.g. the blue qubits implement a constraint $(\hat\sigma_z^{(1)} \hat\sigma_z^{(2)} + \hat\sigma_z^{(2)} \hat\sigma_z^{(3)} + \hat\sigma_z^{(2)} \hat\sigma_z^{(6)})\ket\psi=a\ket\psi$, where $a \in \{-3, -1, 1, 3\}$. The two product constraints $\hat\sigma_z^{(3)} \hat\sigma_z^{(6)}\ket\psi=\ket\psi$ and $\hat\sigma_z^{(1)} \hat\sigma_z^{(2)}\hat\sigma_z^{(7)}\ket\psi=\ket\psi$ are directly incorporated in the parity constraints, represented by colored lines. For example, the condition $\hat\sigma_z^{(3, 6)}=\hat I$ has been used in the black line in the bottom right to implement $ \hat\sigma_z^{(5, 6, 7)} \hat\sigma_z^{(3, 5, 7)}\hat\sigma_z^{(3, 6)}\ket\psi=\hat\sigma_z^{(5, 6, 7)} \hat\sigma_z^{(3, 5, 7)}\ket\psi=\ket\psi$. The curved lines show the two places where bridging is required. 
    }
    \label{fig:constrained_example}
\end{figure}

In Fig.~\ref{fig:constrained_example} we assume a quantum device that can implement such exchange Hamiltonians and show an example of an optimization problem with both product constraints and polynomial constraints. Note that in this example not all connections between qubits are required. In cases where the qubits in a constraint can be connected in multiple ways, this reduced connectivity requirement may make it possible to route around missing links without overhead. 

Due to the restrictions of local polynomial constraints, bridging is very useful when solving these kinds of problems. When optimizing the layout of qubits to make the parity constraints as local as possible, the polynomial constraints should be kept strictly local. This is achieved by only allowing optimization moves that preserve locality of the polynomial constraints.

\section{Conclusion}
In this work, we presented methods to compile constrained optimization problems in the parity architecture, which also allows one to use the parity mapping on devices that are irregularly or sparsely connected. Finding a valid layout mapping where enough parity constraints are present to complete the mapping is now straight-forward, because even non-local constraints can be implemented without SWAP gates. The introduced construction via bridging gives clear rules while providing a lot of freedom in the implementation and thus room for optimization in the required number of gates and circuit depth. For example, solving the problem of finding minimal Steiner trees for arbitrary device connectivity remains a computationally hard problem, but approximate algorithms can already lead to good results. The scaling of the circuit depth with problem and device size will depend on the specific problem and connectivity of the device. Improving the connectivity of the device will allow more local constraints, which in turn allows for more parallelization. The presented approach can thus be used both to improve the circuit implementation on a given device, and to design future devices in a way that keeps the circuit depth manageable for relevant problems. While, in this work, we demonstrated the power of the bridge technique to implement parity constraints, its application goes beyond the parity architecture and can also improve the implementation of many other non-local operators.

\section{Acknowledgements}
The authors would like to thank Valentin Stauber for crucial feedback and discussions about minimal Steiner tree constructions, as well as Michael Fellner for his detailed review and support with quantum circuit transpilers, and Sagar Kale for feedback about the manuscript.
This study was supported by the Austrian Research Promotion Agency (FFG Project No. 884444, QFTE 2020).
Work at the University of Innsbruck was supported by the
  Austrian Science Fund (FWF) through a START grant under Project No. Y1067-N27 and I 6011. 
  This research was funded in whole, or in part, by the Austrian Science Fund (FWF) SFB BeyondC Project No. F7108-N38. For the purpose of open access, the authors have applied a CC BY public copyright licence to any Author Accepted Manuscript version arising from this submission.  
  This project was funded within the
  QuantERA II Programme that has received funding from the European Union's
  Horizon 2020 research and innovation programme under Grant Agreement No.
  101017733. 
%

\end{document}